\documentclass[a4paper,10pt]{article}
\usepackage[cp1251]{inputenc}
\usepackage[english]{babel}
\usepackage{amsmath}
\usepackage{amssymb}
\usepackage{graphicx}
\usepackage{multicol}
\usepackage[pdftex]{color}
\usepackage[square,numbers,comma,sort&compress]{natbib}
\usepackage[pdftex]{hyperref}
\hypersetup{
             final,
             colorlinks=true,
             linkcolor=blue,
             citecolor=red
}

\DeclareMathOperator{\Tr}{Tr}

\begin{document}
{\flushleft\scriptsize{\slshape ISSN 1541-308X, Physics of Wave Phenomena, 2015, Vol. 23, No. 3, pp. 1--6. \textcopyright\,Allerton Press, Inc., 2015.}}

\vspace{0.75cm}

\hrule\vspace{0.07cm}

\hrule

\vspace{1cm}

\begin{center}
\LARGE{\bf Pseudospin Splitting of the Energy Spectrum}

\LARGE{\bf of Planar Polytype Graphene-Based Superlattices}

\vspace{0.75cm}

\large{\bf P. V. Ratnikov and A. P. Silin}

\vspace{0.15cm}

\normalsize

\textit{Lebedev Physical Institute, Russian Academy of Sciences, Leninskii pr. 53, Moscow, 119991 Russia}

\vspace{0.15cm}

e-mail: ratnikov@lpi.ru

\vspace{0.25cm}

Received April 21, 2015
\end{center}

\vspace{0.1cm}
\begin{list}{}
{\rightmargin=1cm\leftmargin=1cm}
\item
\small{\textbf{Abstract---}The energy spectrum of planar polytype graphene-based superlattices has been investigated. It is shown that their energy spectrum undergoes pseudospin splitting due to the asymmetry of quantum wells forming the superlattice potential profile.}

\vspace{0.05cm}

\small{{\bf DOI}: 10.3103/S1541308X15030012}

\end{list}

\vspace{0.5cm}

\normalsize

\begin{multicols}{2}
\begin{center}
1. INTRODUCTION
\end{center}

Currently, researchers pay much attention to graphene-based superlattices. In particular, methods of molecular dynamics were used to calculate graphene-based superlattices with periodically located rows of vacancies \citep{Chern1}. Then superlattices of single-atom thickness, formed by lines of pairs of hydrogen atoms adsorbed on graphene, were calculated within the density functional theory \citep{Chern2}. Rippled graphene, which can be considered as a superlattice with a one-dimensional periodic potential of ripples, was investigated in \citep{Isacsson, Guinea}. Analytical studies were performed on superlattices obtained by applying a periodic electrostatic potential \citep{Bai, Bairbier, Park1, Park2} or periodically arranged magnetic barriers \citep{Masir1, Masir2, Ghosh} to graphene. A detailed review of graphene-based semiconductor heterostructures can be found in \citep{Chern3}.

We investigate planar superlattices based on gapless graphene and its gapped modifications. The main concepts of these superlattices were formulated in \citep{Ratnikov1}, where a dispersion relation for charge carriers in these structures was derived. Then a very simple example of such superlattices was considered: two-type superlattices composed of alternating strips of gapless graphene and its band gap modification.

In \citep{Ratnikov2} we studied graphene-based quantum wells (QWs): planar heterostructures based on graphene, where gapped graphene modifications play a role of potential barriers. In particular, it was shown that the energy spectrum of asymmetric QWs (containing different gapped graphene modifications) is split with respect to pseudospin: the dispersion curves in different valleys do not coincide. This result suggests that pseudospin splitting should also occur in polytype graphene-based superlattices. This splitting is similar in many respects to the spin splitting of the energy spectrum in narrow-gap heterostructures \citep{Kolesnikov, Andryushin}.

Note also that recently we proposed and investigated another version of planar superlattices, which~is peculiar for graphene \citep{Ratnikov3}. Using the Fermi velocity engineering in gapless graphene, one can fabricate structures with periodically modulated Fermi velocity.

In this paper, we discuss the pseudospin splitting of the energy spectrum of a polytype superlattice. As~an~example of polytype graphene-based superlattices, we consider a three-type superlattice in the form $A$---$B$---$C$, where $A$ and $C$ are gapped graphene modifications with different bandgap widths, and $B$ is gapless graphene. This is the simplest example of a polytype superlattice which retains fundamental features of polytype superlattices.

To implement a three-type graphene-based superlattice, we purpose a version involving gapless graphene and its gapped modifications obtained as a result of deposition of a gapless graphene sheet on a particular substrate (e.g., hexagonal boron nitride h-BN) or due to deposition of particular atoms or molecules (for example, CrO$_3$) on the surface of gapless graphene. This version is presented pictorially in \hyperlink{fig1}{Fig. 1}.
%\vspace{0.4cm}

%\begin{figure}[b!]
\begin{center}
\hypertarget{fig1}{}
\includegraphics[width=0.5\textwidth]{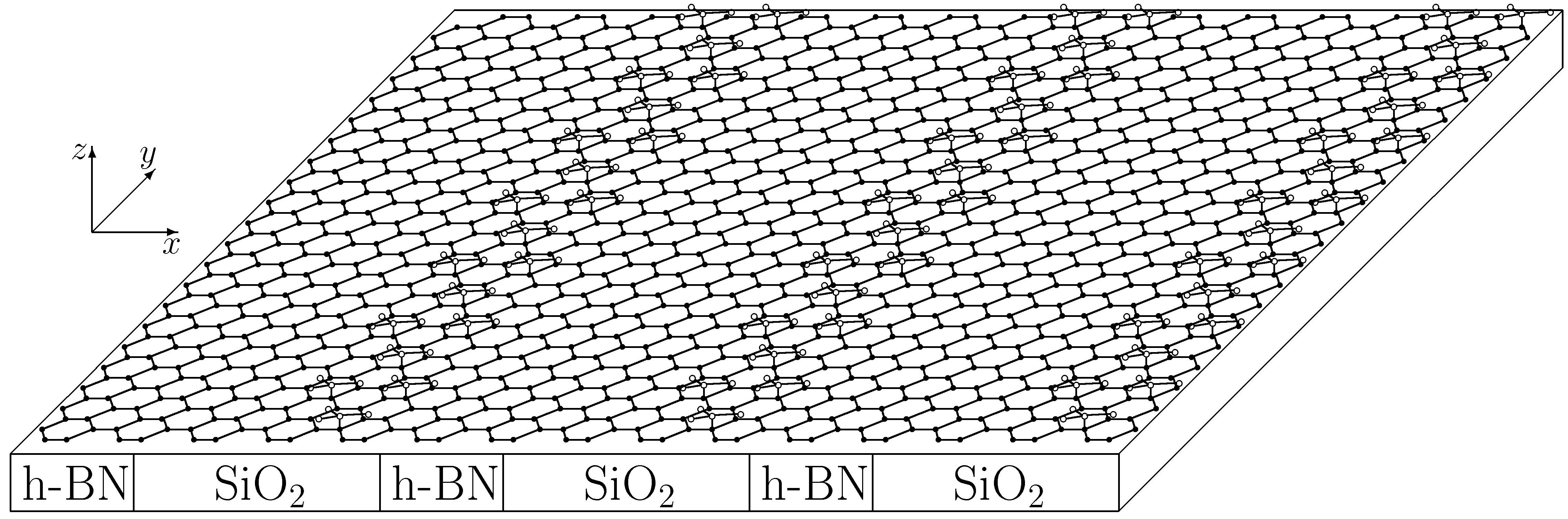}
\end{center}

\hspace{-0.55cm}\small{{\bf Fig. 1.} A version of a three-type superlattice: a graphene sheet on a stripped substrate consisted of SiO$_2$ with deposited stripes of CrO$_3$ molecules and h-BN.}
%\end{figure}

\newpage
\normalsize

\begin{center}
2. MODEL
\end{center}

\vspace{0.15cm}

Let region I be a layer of gapped graphene modification with a bandgap half-width $\Delta_I$ and work function~$V_I$; the thickness of this region is $d_I$. Region II is a layer of gapless graphene with a thickness $d_{II}$ and zero work function. Region III is a layer of gapped graphene modification with a bandgap half-width $\Delta_{III}$, work function $V_{III}$, and thickness $d_{III}$ (\hyperlink{fig2}{Fig. 2}). The superlattice period is $d=d_I+d_{II}+d_{III}$.

\vspace{0.15cm}

Expecting to find pseudospin splitting in the ener-gy spectrum of the superlattice under consideration, we should investigate the solutions to the 4$\times$4 matrix Dirac equation describing both valleys. As was shown in \citep{Ratnikov4}, one can pass to the 2$\times$2 matrix equation for the components of the eigenspinor of pseudo-parity ope-rator. Note that this equation contains explicitly the~operator eigenvalue --- pseudo-parity $\lambda$, which distinguishes states from different valleys: $\lambda=+1$ for the~states near the $K$ point and $\lambda=-1$ for the states near the $K^\prime$ point. We will use specifically this equation to make the computation less cumbersome:

\begin{equation}\label{1}
\left(v_F\sigma_x\widehat{p}_x+\lambda v_F\sigma_yk_y+\Delta\sigma_z+V\right)\psi_\lambda(x)=E_\lambda\psi_\lambda(x),
\end{equation}
\vspace{0.1cm}

\hspace{-0.55cm}where $v_F$ is the Fermi velocity (it is constant in all superlattice regions); $\sigma_x$, $\sigma_y$, and $\sigma_z$ are the Pauli matrices, which act in the sublattice space; $\widehat{p}_x=-i\partial_x$ is the $x$ component of the momentum operator (from here on, Planck's constant is assumed to be unity); $k_y$~is~the~$y$ component of the quasimomentum (charge carriers move freely along the $y$ axis); and $\Delta$ and $V$ are quantities periodically changing along the $x$ axis with a period $d$,

\vspace{0.25cm}
%\begin{figure}[b!]
\begin{center}
\hypertarget{fig2}{}
\includegraphics[width=0.5\textwidth]{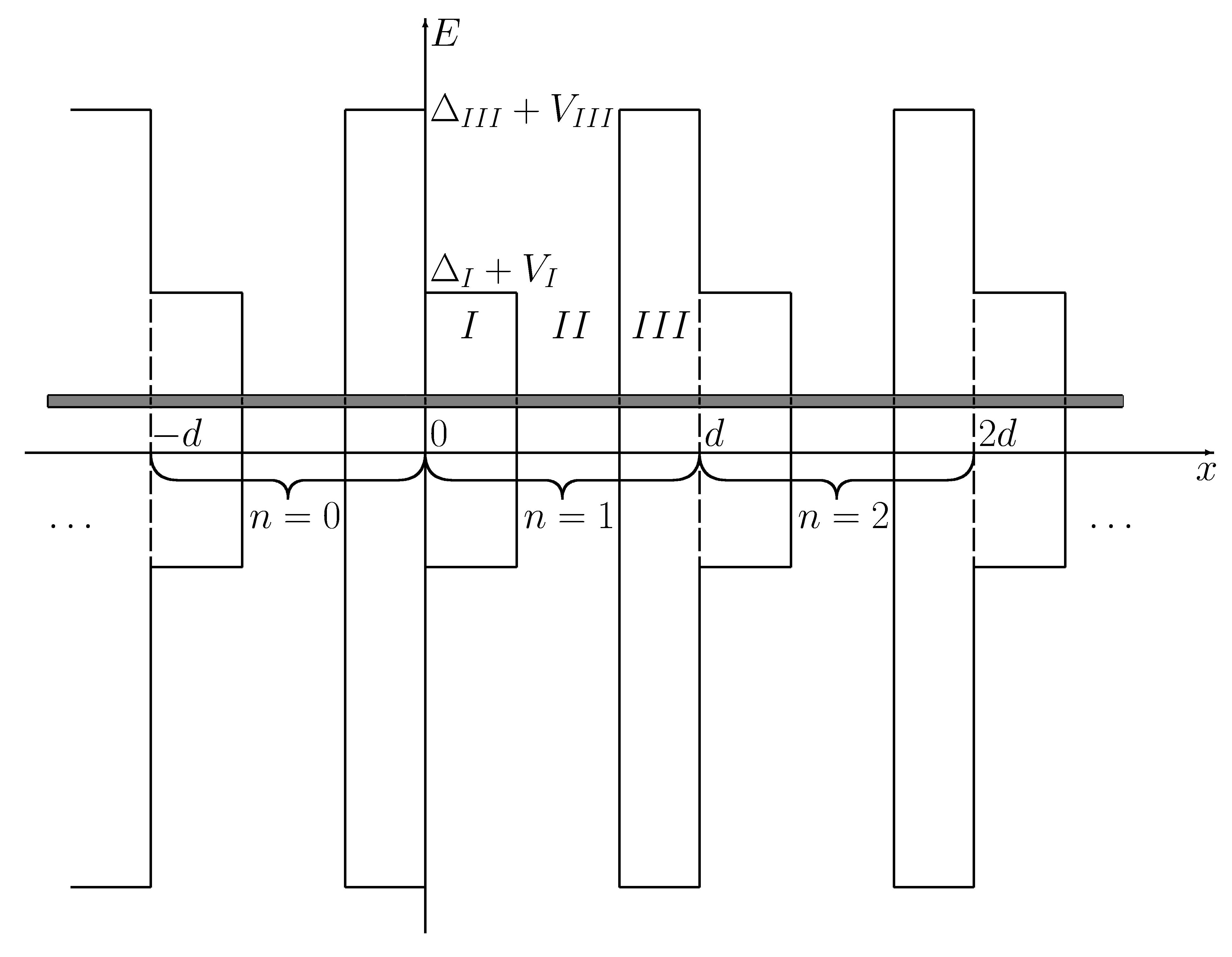}
\end{center}

\hspace{-0.55cm}\small{{\bf Fig. 2.} Energy band diagram of a three-type superlattice, supercell of which consists of a gapless graphene strip and two strips of its gapped modifications. The position of one the lower minibands of interest is shown by a gray strip.}
%\end{figure}

\normalsize
\begin{equation*}
\begin{split}
\Delta=\begin{cases}\Delta_I,& (n-1)d<x<(n-1)d+d_I,\\
0,& (n-1)d+d_I<x<nd-d_{III},\\
\Delta_{III},& nd-d_{III}<x<nd;\end{cases}\\
V=\begin{cases}V_I,& (n-1)d<x<(n-1)d+d_I,\\
0,& (n-1)d+d_I<x<nd-d_{III},\\
V_{III},& nd-d_{III}<x<nd,\end{cases}
\end{split}
\end{equation*}
where $n$ is an integer enumerating supercells (see \hyperlink{fig2}{Fig.~2}).

Let us write solutions to Eq. \eqref{1} in each region within the $n$th supercell:

~

(i) in region I ($0<x<d_I$),
\begin{equation}\label{2}
\psi^{(1)}_{\lambda n}(x)=\Omega^{(1)}_{\lambda k_1}\begin{pmatrix}a^{(1)}_{\lambda n}\\ c^{(1)}_{\lambda n}\end{pmatrix},
\end{equation}
where
\begin{equation*}
\begin{split}
&\Omega^{(1)}_{\lambda k_1}=A\begin{pmatrix}1&1\\ \alpha^{(+)}_\lambda&-\alpha^{(-)}_\lambda\end{pmatrix}\exp(-k_1x\sigma_z),\\
&\alpha^{(\pm)}_\lambda=\frac{iv_F(k_1\pm\lambda k_y)}{E_\lambda+\Delta_I-V_I};
\end{split}
\end{equation*}

~

(ii) in region II ($d_I<x<d_I+d_{II}$),
\begin{equation}\label{3}
\psi^{(2)}_{\lambda n}(x)=\Omega^{(2)}_{\lambda k_2}\begin{pmatrix}a^{(2)}_{\lambda n}\\ c^{(2)}_{\lambda n}\end{pmatrix},
\end{equation}
where
\begin{equation*}
\begin{split}
&\Omega^{(2)}_{\lambda k_2}=A\begin{pmatrix}1&1\\ \beta^{(+)}_\lambda&-\beta^{(-)}_\lambda\end{pmatrix}\exp(ik_2x\sigma_z),\\
&\beta^{(\pm)}_\lambda=\frac{v_F(k_2\pm i\lambda k_y)}{E_\lambda};
\end{split}
\end{equation*}

~

(iii) in region III ($d_I+d_{II}<x<d$)
\begin{equation}\label{4}
\psi^{(3)}_{\lambda n}(x)=\Omega^{(3)}_{\lambda k_3}\begin{pmatrix}a^{(3)}_{\lambda n}\\ c^{(3)}_{\lambda n}\end{pmatrix},
\end{equation}
where
\begin{equation*}
\begin{split}
&\Omega^{(3)}_{\lambda k_3}=A\begin{pmatrix}1&1\\ \gamma^{(+)}_\lambda&-\gamma^{(-)}_\lambda\end{pmatrix}\exp(-k_3x\sigma_z),\\
&\gamma^{(\pm)}_\lambda=\frac{iv_F(k_3\pm\lambda k_y)}{E_\lambda+\Delta_{III}-V_{III}}.
\end{split}
\end{equation*}

$A$ is the normalization factor for all three regions. The $k_1$, $k_2$, and $k_3$ values are related by the correspon-ding expressions to energy $E_\lambda$:
\begin{equation}
\begin{split}
&v_F^2k^2_{1,3}=\Delta^2_{I,III}-(E_\lambda-V_{I,III})^2+v_F^2k^2_y,\\
&v_F^2k^2_2=E^2_\lambda-v_F^2k^2_y.
\end{split}
\end{equation}
The $a^{(j)}_{\lambda n}$ and $c^{(j)}_{\lambda n}$ values are constants, which can be found from the boundary conditions.

Now, based on solutions \eqref{2}--\eqref{4}, we can construct the wave function of charge carriers in the superlattice. In this study, we are not interested in the interfacial states (see, for example, \citep{Ratnikov4}); therefore, $v_F$ is assumed to be the same in all regions. In this case, the boundary conditions for the superlattice include the requirement for the wave function continuity within each supercell and between neighboring supercells \citep{Ratnikov2, Silin}:

~

(a) continuity of the wave function between regions I and II at the boundary $x=d_I$
\begin{equation}\label{6}
\Omega^{(1)}_{\lambda k_1}(d_I)\begin{pmatrix}a^{(1)}_{\lambda n}\\ c^{(1)}_{\lambda n}\end{pmatrix}=\Omega^{(2)}_{\lambda k_2}(d_I)\begin{pmatrix}a^{(2)}_{\lambda n}\\ c^{(2)}_{\lambda n}\end{pmatrix},
\end{equation}

~

(b) continuity of the wave function between regions II and III at the boundary $x=d_{II}$
\begin{equation}\label{7}
\Omega^{(2)}_{\lambda k_2}(d_I+d_{II})\begin{pmatrix}a^{(2)}_{\lambda n}\\ c^{(2)}_{\lambda n}\end{pmatrix}=\Omega^{(3)}_{\lambda k_3}(d_I+d_{II})\begin{pmatrix}a^{(3)}_{\lambda n}\\ c^{(3)}_{\lambda n}\end{pmatrix},
\end{equation}

~

(c) continuity of the wave function at the boundary between region III $n$th supercell and the region I of the $(n+1)$th supercell
\begin{equation}\label{8}
\Omega^{(3)}_{\lambda k_3}(d)\begin{pmatrix}a^{(3)}_{\lambda n}\\ c^{(3)}_{\lambda n}\end{pmatrix}=\Omega^{(1)}_{\lambda k_1}(0)\begin{pmatrix}a^{(1)}_{\lambda n+1}\\ c^{(1)}_{\lambda n+1}\end{pmatrix}.
\end{equation}

In view of the periodicity of the system, the Bloch conditions
\begin{equation}
\psi^{(j)}_{\lambda n}(x+d)=\psi^{(j)}_{\lambda n}(x)\exp(ik_xd)
\end{equation}
must also be satisfied for all three regions ($j=1,\,2,\,3$).

\vspace{0.5cm}
\hrule

\begin{center}
3. TRANSFER MATRIX\\ AND DISPERSION RELATION
\end{center}

Furthermore, we will determine the transfer matrix ($T$ matrix) that links coefficients of the wave function in neighboring supercells \citep{Ratnikov1, Ratnikov3}:
\begin{equation}\label{10}
\begin{pmatrix}a^{(j)}_{\lambda n+1}\\ c^{(j)}_{\lambda n+1}\end{pmatrix}=T^{(j)}_\lambda\begin{pmatrix}a^{(j)}_{\lambda n}\\ c^{(j)}_{\lambda n}\end{pmatrix}.
\end{equation}

Definition \eqref{10} shows that there may be three versions of the $T$ matrix linking the corresponding pairs of constants from solutions \eqref{2}--\eqref{4} for regions I, II, or III. This arbitrariness in definition is allowable, because all three versions are interrelated by cyclic permutation of $\Omega$ matrices, and the dispersion relation includes the $T$-matrix trace. In particular, using equalities \eqref{6}--\eqref{8}, one can easily derive the expression
\begin{equation}\label{11}
\begin{split}
T^{(1)}_\lambda&=\left(\Omega^{(1)}_{\lambda k_1}(0)\right)^{-1}\Omega^{(3)}_{\lambda k_3}(d)\left(\Omega^{(3)}_{\lambda k_3}(d_I+d_{II})\right)^{-1}\\
&\times\Omega^{(2)}_{\lambda k_2}(d_I+d_{II})\left(\Omega^{(2)}_{\lambda k_2}(d_I)\right)^{-1}\Omega^{(1)}_{\lambda k_1}(d_I).
\end{split}
\end{equation}

The dispersion relation has the form \citep{Ratnikov1, Ratnikov3}
\begin{equation}\label{12}
\Tr T^{(j)}_\lambda=2\cos(k_xd).
\end{equation}
Having calculated the $T$-matrix trace, we obtain the dispersion relation in the form
\end{multicols}
\begin{equation}\label{13}
\begin{split}
&\left\{q^{(-)}_1\left[g^{(+)}_1\cos(k_2d_{II})+f^{(+)}_1\sin(k_2d_{II})\right]\exp(-k_3d_{III})\right.\\
&\left.-q^{(-)}_2\left[g^{(+)}_2\cos(k_2d_{II})+f^{(+)}_2\sin(k_2d_{II})\right]\exp(k_3d_{III})\right\}\exp(-k_1d_I)\\
&+\left\{q^{(+)}_1\left[g^{(-)}_1\cos(k_2d_{II})-f^{(-)}_1\sin(k_2d_{II})\right]\exp(k_3d_{III})\right.\\
&\left.-q^{(+)}_2\left[g^{(-)}_2\cos(k_2d_{II})-f^{(-)}_2\sin(k_2d_{II})\right]\exp(-k_3d_{III})\right\}\exp(k_1d_I)=\\
&=\frac{8v^3_Fk_1k_2k_3}{E_\lambda\left(E_\lambda+\Delta_I-V_I\right)\left(E_\lambda+\Delta_{III}-V_{III}\right)}\cos(k_xd),
\end{split}
\end{equation}
where the following designations are introduced:
\begin{equation*}
\begin{split}
&q^{(\pm)}_1=\frac{v_F(k_1\pm\lambda k_y)}{E_\lambda+\Delta_I-V_I}+\frac{v_F(k_3\mp\lambda k_y)}{E_\lambda+\Delta_{III}-V_{III}},\hspace{0.15cm}q^{(\pm)}_2=\frac{v_F(k_1\pm\lambda k_y)}{E_\lambda+\Delta_I-V_I}-\frac{v_F(k_3\pm\lambda k_y)}{E_\lambda+\Delta_{III}-V_{III}},\\
&g^{(\pm)}_1=\frac{v_Fk_2}{E_\lambda}q^{(\pm)}_1,\hspace{0.15cm}g^{(\pm)}_2=\frac{v_Fk_2}{E_\lambda}q^{(\pm)}_2,\\
&f^{(\pm)}_1=1\mp\lambda k_y\frac{v_F}{E_\lambda}\left(\frac{v_F(k_1\pm\lambda k_y)}{E_\lambda+\Delta_I-V_I}-\frac{v_F(k_3\mp\lambda k_y)}{E_\lambda+\Delta_{III}-V_{III}}\right)-
\frac{v^2_F(k_1\pm\lambda k_y)(k_3\mp\lambda k_y)}{(E_\lambda+\Delta_I-V_I)(E_\lambda+\Delta_{III}-V_{III})},\\
&f^{(\pm)}_2=1\mp\lambda k_y\frac{v_F}{E_\lambda}\left(\frac{v_F(k_1\pm\lambda k_y)}{E_\lambda+\Delta_I-V_I}+\frac{v_F(k_3\pm\lambda k_y)}{E_\lambda+\Delta_{III}-V_{III}}\right)+
\frac{v^2_F(k_1\pm\lambda k_y)(k_3\pm\lambda k_y)}{(E_\lambda+\Delta_I-V_I)(E_\lambda+\Delta_{III}-V_{III})}.
\end{split}
\end{equation*}

An analysis of Eq. \eqref{13} shows that the necessary condition for pseudospin splitting of the energy spectrum is the absence of electron--hole symmetry. The symmetry of the system in the energy space with respect the replacement $E\rightarrow-E$ should be absent, and $V_I$ or $V_{III}$ should differ from zero. A sufficient condition for pseudospin splitting of the energy spectrum of the superlattice is the inequality
\begin{equation}\label{14}
\frac{\Delta_I}{V_I}\neq\frac{\Delta_{III}}{V_{III}}.
\end{equation}

\begin{multicols}{2}
Dispersion relation \eqref{13} can be simplified in the case where QWs exist for only electrons or for only holes~\citep{Ratnikov1}. In the case of electrons, one must assume that $V_I=\Delta_I$ and $V_{III}=\Delta_{III}$; then relation \eqref{13} yields
\begin{equation}\label{15}
\begin{split}
&\cosh(k_1d_I)\cos(k_2d_{II})\cosh(k_3d_{III})\\
&+\frac{1}{2}\left[X^{(+)}_{1,3}\sinh(k_1d_I)\cos(k_2d_{II})\sinh(k_3d_{III})\right.\\
&+X^{(-)}_{1,2}\sinh(k_1d_I)\sin(k_2d_{II})\cosh(k_3d_{III})\\
&\left.+X^{(-)}_{3,2}\cosh(k_1d_I)\sin(k_2d_{II})\sinh(k_3d_{III})\right]=\cos(k_xd),
\end{split}
\end{equation}
where $X^{(\pm)}_{i,j}=x_{i,j}\pm x^{-1}_{i,j}$ and $x_{i,j}=k_i/k_j$.\, After the analytical continuation $k_1\rightarrow ik_1$ and $k_3\rightarrow ik_3$, Eq. \eqref{15} coincides with the well-known nonrelativistic equation (see, for example, monograph \citep{Herman}). Inequa-lity \eqref{14} is not fulfilled, and pseudospin splitting of the~energy spectrum is absent.

The limiting transition from superlattice to individual QW occurs if the potential barriers are sufficiently wide. Under these conditions, $k_1d_I$ and $k_3d_{III}$ are much larger than unity, and the dispersion relation for superlattice \eqref{13} yields
\begin{equation}\label{16}
\tan(k_2d_{II})=\frac{g^{(-)}_1}{f^{(-)}_1}
\end{equation}
for an individual QW, the pseudospin splitting in which was considered in \citep{Ratnikov2}.

\begin{center}
4. NUMERICAL CALCULATIONS
\end{center}

We performed calculations for the aforementioned gapped graphene modifications: graphene deposited on h-BN strips ($\Delta_I=26.5$ meV) and graphene with deposited CrO$_3$ molecules ($\Delta_{III}=60$ meV). The strip widths were taken to be $d_I=d_{III}=21.3$ nm and $d_{II}=42.6$ nm.

For\,\, simplicity,\,\, we\,\, assume\,\, that\,\, $V_I=0$.\,\, The exact value of the work function for the second gapped graphene modification is unknown; at the same time, according to the energy band calculations, it is negative \citep{Zanella}. We chose $V_{III}$ to be --10 meV.

The calculation results for the lower electron miniband are presented in \hyperlink{fig3}{Fig. 3}. One can see that the electron valleys for different pseudospins ($\lambda=\pm1$) un-

%\begin{figure}[b!]
\begin{center}
\hypertarget{fig3}{}
\includegraphics[width=0.5\textwidth]{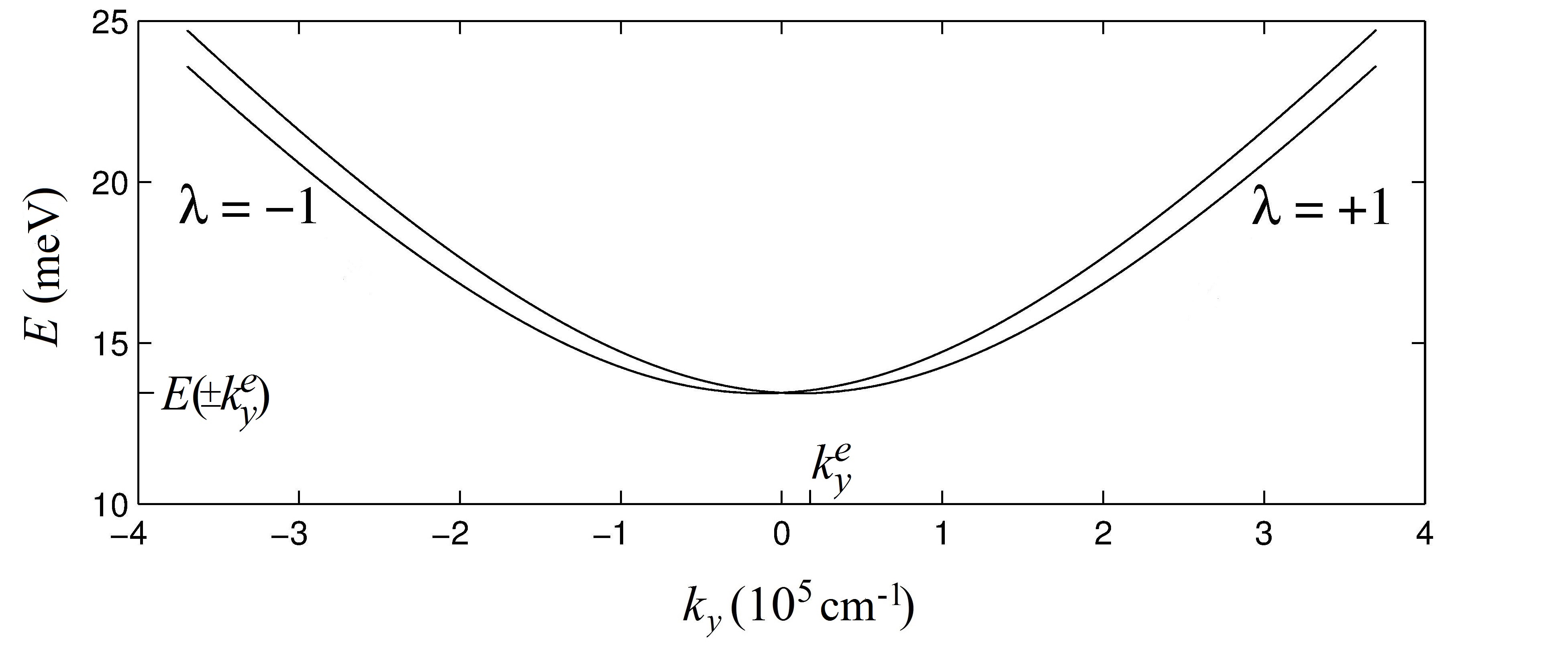}
\end{center}

\hspace{-0.55cm}\small{{\bf Fig. 3.} Results of numerical calculation for a lower electron miniband in a three-type graphene-based superlattice. The quasi-momentum component $k_x=0$.}
%\end{figure}

\normalsize

\hspace{-0.55cm}derwent splitting. The extremum of valley is at the point $\lambda k^e_y$. For the superlattice under consideration, we~have the following parameters: $k^e_y\approx1.9\times10^4$ cm$^{-1}$, $E^e_\lambda(\lambda k^e_y)\approx13.5$ meV, and energy splitting $\Delta^e_{ps}=E^e_{\lambda=-1}(k^e_\lambda)-E^e_{\lambda=+1}(k^e_\lambda)\approx0.06$ meV for the lower electron miniband and $k^h_y\approx1.7\times10^4$ cm$^{-1}$, $E^h_\lambda(\lambda k^h_y)\approx-15.5$ meV, and $\Delta^h_{ps}=E^h_{\lambda=+1}(k^h_\lambda)-E^h_{\lambda=-1}(k^h_\lambda)\approx0.05$~meV for the upper hole miniband. The thus formed energy gap is $E_G=E^e_\lambda(k^e_y)-E^h_\lambda(k^h_y)\approx29$~meV.

\begin{center}
5. CONCLUSIONS
\end{center}

We proposed a new class of systems based on graphene: planar polytype superlattices composed of gapless graphene and its different gapped modifications. These superlattices, formed by multiply repeated asymmetric QWs, are characterized by pseudospin splitting of the energy spectrum. We investigated the conditions for occurrence of this splitting.

The spectrum calculated by us is similar to that considered by Bychkov and Rashba \citep{Bychkov}. The relatioship between our Hamiltonian and the Bychkov--Rashba Hamiltonian was considered by us in detail~in~\citep{Ratnikov2}.

The results obtained barely change if we consider smooth heterojunctions instead of sharp ones \citep{Ratnikov2}.

Due to the pseudospin splitting of the energy spectrum, the bandgap in this superlattice becomes indirect (in contrast to the superlattice considered in \citep{Ratnikov1}). The superlattice analyzed by us is similar to a narrow-gap indirect semiconductor and can be used in IR optics. The characteristic bandgap width is $E_G\cong30$~meV. The spacing between the extrema of electron and hole valleys with identical $\lambda$ is $\widetilde{k}_y=k^e_y+k^h_y\cong4\times10^4$ cm$^{-1}$. The desired parameters of superlattice spectrum ($E_G$ and $\widetilde{k}_y$) can be obtained by changing the superlattice characteristics ($d_{I,II,III}$, $V_{I,III}$, and $\Delta_{I,III}$).

\end{multicols}
\end{document}